\documentclass[letterpaper,prl,twocolumn,amsmath,amssymb,groupedaddress,bibliography]{revtex4-1}
\usepackage[utf8]{inputenc}
\usepackage[draft]{hyperref}
\usepackage{graphicx}
\usepackage{dcolumn}
\usepackage{bm}
\usepackage{amssymb}
\usepackage{physics}
\usepackage{color}
\usepackage{soul}
\usepackage{epstopdf}
\usepackage{float}

\begin{document}

\title {Raman Laser Switching Induced by Cascaded Light Scattering}
\author{Sho Kasumie,$^1$  Fuchuan Lei,$^{1,}$ }
\email{fuchuan.lei@oist.jp} 
\author{Jonathan M Ward,$^1$ Xuefeng Jiang,$^2$ Lan Yang,$^2$ and S\'ile Nic Chormaic$^{1,3}$}
\email{sile.nicchormaic@oist.jp} 

\affiliation{$^1$Light-Matter Interactions Unit, Okinawa Institute of Science and Technology Graduate University, Onna, Okinawa 904-0495, Japan\\
$^2$Department of Electrical and Systems Engineering, Washington University, St. Louis, Missouri 63130, USA\\
$^3$Universit\'e Grenoble Alpes, CNRS, Grenoble INP, Institut N\'eel, 38000 Grenoble, France}


\date{\today}

\begin{abstract}
We show that, in multimode Raman lasers, cascaded light scattering (CLS) not only extends the optical frequency range, but could also modulate the laser dynamics. The origin of this phenomenon is based on the fact that many Raman lasing modes are directly correlated through CLS. The coupled-mode equations only describe single-mode cascaded Raman lasers and are insufficient for describing the multimode case. In this work, we introduce additional terms to account for intermodal interaction and, thence, reveal the physical mechanism behind the mode-switching phenomenon. Additionally, mode-switching controlled solely by a single-mode pump in whispering gallery mode (WGM) silica Raman lasers is demonstrated. As the intracavity pump power is increased, laser switching happens between two adjacent WGMs in the same mode family.
\end{abstract}

\pacs{Valid PACS appear here}
\maketitle



\begin{figure}[h]
\centering
\includegraphics[width=\linewidth]{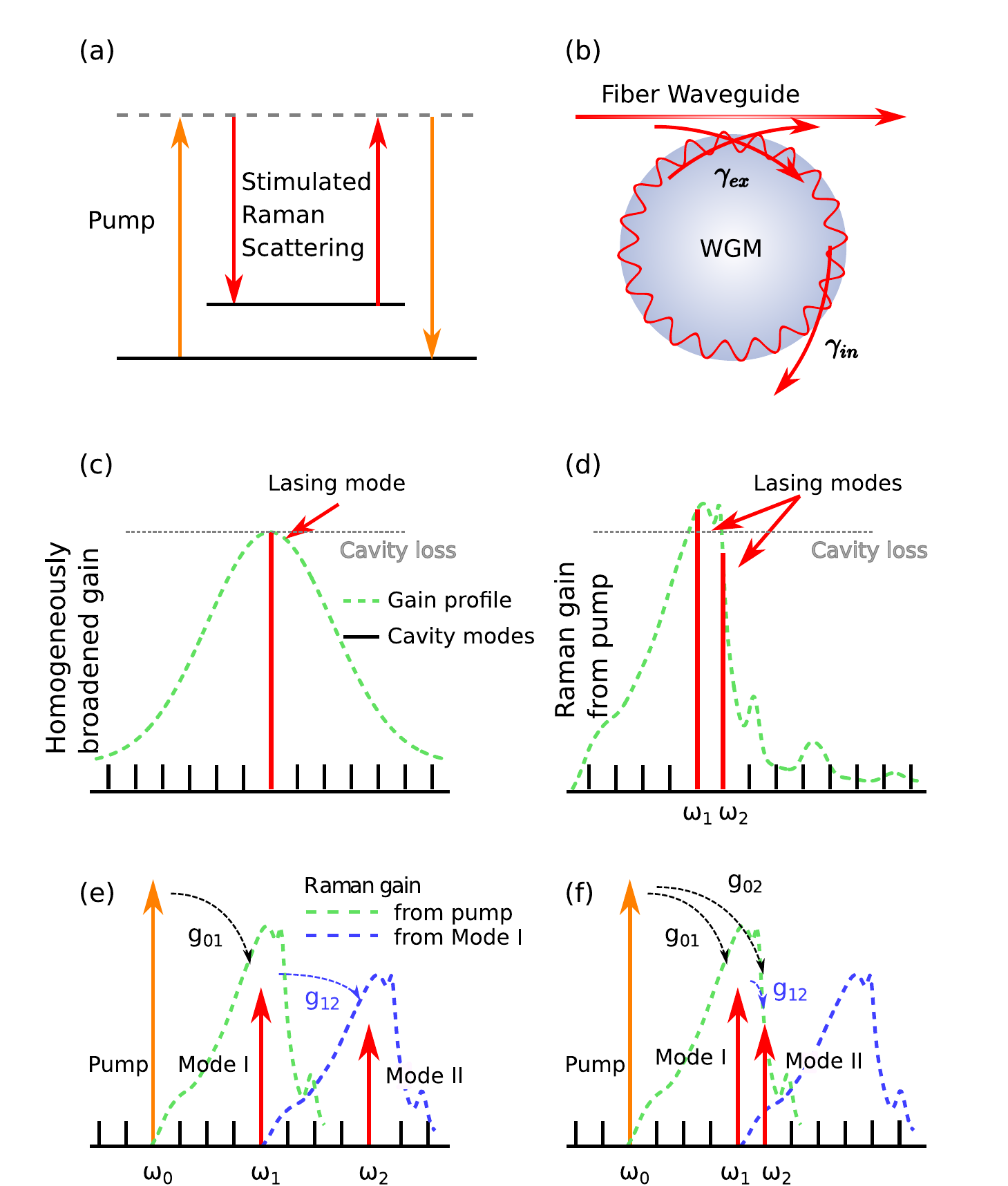}
\caption{(a) Schematic diagram of the stimulated Raman scattering process. (b) Schematic of the experimental system. The input laser is coupled to a silica microsphere resonator through a tapered fiber waveguide, with a coupling constant,  $\gamma_{ex}$, and  resonator intrinsic loss,  $\gamma_{in}$.  Comparison of the gain spectrum and the lasing modes between (c) a typical homogeneously broadened laser and (d) a silica Raman laser. Schematic of (e) a cascaded  Raman laser in single-mode fashion and (f) a two-mode Raman laser.}
\label{fig:RamanDiag}
\end{figure}

Stimulated light scattering, such as stimulated Raman scattering (SRS) and stimulated Brillouin scattering (SBS), can lead to the generation of coherent photons in different materials and geometries \cite{boyd2003nonlinear,lin2017nonlinear}. Unlike  conventional inversion lasers, the emissions of stimulated-scattering lasers are not limited to specific wavebands because no real energy levels are required; this provides unique advantages in many applications, e.g. arbitrary wavelength conversion \cite{Spillane2002,boyraz2004demonstration,rong2005all}, high-quality microwave generation \cite{li2013microwave},  gyroscopes \cite{li2017microresonator}, sensing \cite{jiang2018whispering,Ozdemira2014,jiang2013free,zhao2017raman,li2014single,chen2017tunable},  mode control \cite{yang2015raman,wen2012all}, etc. Considering that the stimulated gain arises from coherent coupling between a pump field and a Stokes field mediated by another field, e.g. a phonon field, the newly generated Stokes field may also act as a pump for the generation of  further Stokes fields - an effect known as  cascaded light scattering (CLS), resulting in  cascaded Raman and  Brillouin lasers  \cite{Min2003,kippenberg2004theoretical,rong2008cascaded,grudinin2009brillouin, lin2014cascaded}. Due to the presence of CLS,  stimulated lasers are usually multimode when a high pump power is applied. Nevertheless, it is generally assumed that the dynamics of stimulated lasers involving CLS is trivial and usually  attention is focused on the interactions of pump and Stokes waves. This is indeed true for Brillouin lasers because of the phase-matching condition \cite{grudinin2009brillouin,tomes2009photonic}. However, in Raman lasers, as the phase matching condition is automatically satisfied \cite{Spillane2002}, see Fig.\ref{fig:RamanDiag}(a)\cite{boyd2003nonlinear}, CLS could lead to many lasing modes being coupled together, thence rendering the behavior of such lasers to be quite unique, especially for multimode lasing \cite{ge2016interaction,leymann2017pump,zhang2019all}. As a specific example, in this Letter, we show both theoretically and experimentally, that  CLS can induce mode-switching in multimode Raman lasers. This study suggests the importance of considering CLS effects for Raman lasing in dielectric resonators.

In this work, a whispering gallery mode (WGM) silica Raman laser is chosen as the experimental platform \cite{ward2011wgm}. As depicted in Fig. \ref{fig:RamanDiag}(b), the light fields can be coupled into and out of the WGM resonator through a waveguide. The Stokes light (i.e. the Raman laser) can be generated in a WGM resonator as long as the following requirements are met: (i) the frequency of the Stokes light coincides with a high Q WGM and (ii) the Stokes mode has sufficient spatial overlap with the pump mode, i.e., mode  overlap. 
With a single-mode pump, the Raman gain of a silica matrix can be considered to be homogeneously broadened if one neglects CLS,  see Fig. \ref{fig:RamanDiag}(c). Under this circumstance, only the mode with the highest gain can lase, even if all the modes have the same losses, i.e., Q factors, since the gain is clamped at the lasing threshold \cite{verdeyen1989laser}.
At this point, a question would naturally arise, how can one explain the frequent occurrence of multimode Raman lasing in a single-mode pumped microresonator?

Here, we show that, due to the existence of  CLS, two (or multiple) modes with unequal gain provided by the pump, but equal cavity loss (more generally, unequal gain/cavity loss ratio) lase simultaneously, see Fig. \ref{fig:RamanDiag}(d). It is noteworthy to point out that this CLS-assisted, multimode lasing scenario differs from  conventional cascaded single-mode Raman lasing , where a single mode is generated for each order of lasing, as shown in Fig. \ref{fig:RamanDiag}(e). The first order Raman lasing mode (Mode \uppercase\expandafter{\romannumeral1}), originating from the pump, could generate the subsequent Stokes field, i.e., the second order Raman laser (Mode \uppercase\expandafter{\romannumeral2}), and the typical frequency shift between the pump and Stokes is about 14 THz. In this case, the pump does not provide gain directly to the second (or even higher order) Stokes fields. Hence, the power of this Raman laser usually decreases with increasing order, and it is impossible to obtain a higher order Raman signal without the presence of its previous order.    The dynamics of such a typical cascaded laser can be simply treated as an cascaded energy transfer process, therefore it is trivial and does not differ much from the conventional single-mode lasing case. However, for the multimode lasing case, the presence of CLS can modulate the laser dynamics significantly. As illustrated in Fig. \ref{fig:RamanDiag}(f), two adjacent modes \textit{in the same mode family} are excited simultaneously by a single-mode pump. Both Raman lasing modes derive their gain from the pump through SRS, but, at the same time, the first Raman mode interacts with the second due to CLS . Though the interaction may be weak, it is not negligible and can account for the existence of multimode Raman lasing, and even play a subtle role in  Raman laser  switching. 

Indeed, the CLS induced coupling between the two lasing modes, seems not so obvious and is easily overlooked. In textbooks, the SRS is usually illustrated with two light fields and a simplified two level system, as shown in Fig. \ref{fig:RamanDiag}(a). However, in real solid materials, like silica \cite{Hollenbeck2002}, silicon \cite{uchinokura1972raman} or silicon carbide \cite{PhysRevB.6.498}, the Raman gain profile is not composed of only several discrete peaks, instead, it is a continuum; therefore, it is natural to introduce a coupling term between the two lasing modes.

In order to understand the multimode Raman laser (see Fig. \ref{fig:RamanDiag}(f)), the standard coupled-mode equations are modified \cite{Min2003} to inclulde $all$ the CLS terms. Given a certain $j$th cavity mode, it couples to the external driving pump in the waveguide via the overlapping of the evanescent fields, and to $all$ the other cavity modes through SRS.  We introduce a summation term into the coupled-mode equations to represent the interactions of the cavity modes. 
The motion of the intracavity field, $E_j$, can now be described as
\begin{equation}
\begin{split}
\dot{E_j}&=\left(-\frac{\gamma_{j}}{2}+\sum_{i<j}g_{ij}|E_i|^2-\sum_{k>j}g_{jk}\frac{\omega_j}{\omega_k}|E_k|^2\right)E_j\\
&\quad +\sqrt[]{\gamma_{ex,0}}\delta_{0,j}s,
\end{split}
\label{eq:generalRaman}
\end{equation}
 where $i$, $j$, and $k$ are mode order indices and the resonant frequencies decrease with the order (the order of the pump mode is set to be 0).   $\gamma_{j}$ and $\gamma_{ex,j}$ denote the total energy decay rate and the extrinsic decay rate into the waveguide, respectively. $\omega_j$ is the resonant frequency, $I_0=|s|^2$ is the input pump power from the waveguide, and $\delta$ is the Kronecker delta function, where $\delta_{ij}=1$ if and only if $i=j$. Here, $g_{ij}$ is the intracavity Raman gain coefficient between mode $i$ and $j$, proportional to the Raman gain spectrum of the bulk silica, $g_R(|\omega_i-\omega_j|)$ \cite{Hollenbeck2002,Min2003}. It is noted that the Raman gain spectrum will not be modified due to the effect of cavity quantum electrodynamics (cQED) \cite{matsko2003cavity}. The second  and  third terms in Eq. 1 describe the gain and the loss caused by the SRS from the  higher and lower frequency modes, respectively. It is important to point out that the  coherent  anti-Stokes Raman scattering is not included in the model since the phase-matching condition is not easily satisfied, as discussed later.

\begin{figure}[h]
\centering
\includegraphics[width=\linewidth]{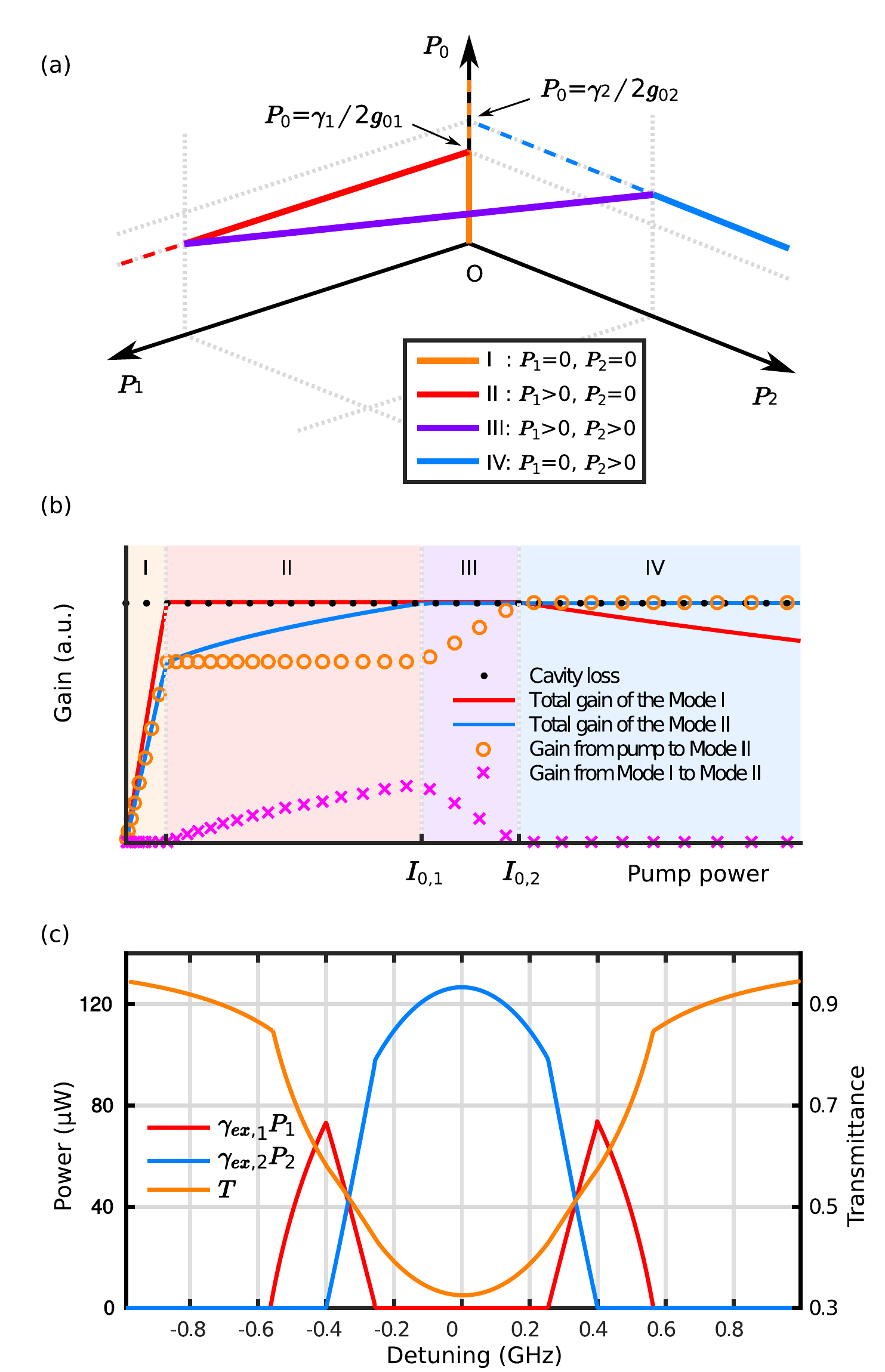}
\caption{ Simulation results based on Eq. (\ref{eq:generalRaman}). (a) $P_0$, $P_1$, and $P_2$ represent the intracavity powers of the pump, the first, and the second Raman fields. The external pump power is increased when detuning is set to 0. Solid lines represent the stable, steady state, while the dashed lines correspond to unstable cases. (b) The total gain of the two Raman modes and gain fraction from pump and Mode \uppercase\expandafter{\romannumeral1} to Mode \uppercase\expandafter{\romannumeral2}. (c) The output of two Raman lasers and the transmission spectrum (T) of the pump mode as it evolves with pump detuning. The parameters used in (b) and (c) are: $\gamma_{j} = 25.3$ $\mu s^{-1}$ and $\gamma_{ex,j} = 26.2$ $\mu s^{-1}$ for all $j = 0,1,2$; the input power in (c) is fixed as $I_0= 0.6$ mW; the angular frequencies are $\omega_0 = 2\pi\times390.6$ THz, $\omega_1 = 2\pi\times377$ THz and $\omega_2 = 2\pi\times375.3$ THz; 
and the intracavity Raman gain coefficients are $g_{01} = 3.6\times10^{18}$ $s^{-1}J^{-1}$, $g_{02} = 2.7\times10^{18}$ $s^{-1}J^{-1}$ and $g_{12} = 0.5\times10^{18}$ $s^{-1}J^{-1}$, estimated based on ref \cite{Hollenbeck2002}.}
\label{fig:RamanSwitch}
\end{figure}

In order to gain insight into this phenomenon,  we consider the simplest case, i.e., one pump mode and two Raman lasing modes (see Fig. \ref{fig:RamanDiag}(f)). The intracavity powers for  steady-state are governed by 
\begin{eqnarray}
P_0\left(-\frac{\gamma_{0}}{2}-g_{01}\frac{\omega_{0}}{\omega_{1}}P_1-g_{02}\frac{\omega_{0}}{\omega_{2}}P_2\right)^2&=&\gamma_{ex,0}I_0,\\
P_1\left(-\frac{\gamma_{1}}{2}+g_{01}P_0-g_{12}\frac{\omega_{1}}{\omega_{2}}P_2\right)&=&0,\\
P_2\left(-\frac{\gamma_{2}}{2}+g_{02}P_0+g_{12}P_1\right)&=&0,\label{eq:P2}
\end{eqnarray}
where $P_0$, $P_1$, and $P_2$ represent the intracavity power for the pump, the first (Mode \uppercase\expandafter{\romannumeral1}), and the second lasing mode (Mode \uppercase\expandafter{\romannumeral2}), respectively.  There are four different steady-state regimes as $I_0$ is increased, see Fig. \ref{fig:RamanSwitch}(a)  and Fig. \ref{fig:RamanSwitch}(b). Mode \uppercase\expandafter{\romannumeral1} is  located in the vicinity of the peak of the Raman gain profile, while Mode \uppercase\expandafter{\romannumeral2} has a lower resonant frequency and a relatively lower Raman gain coefficient. Therefore,  Mode \uppercase\expandafter{\romannumeral1} lases first, as long as its gain can overcome its loss (regime \uppercase\expandafter{\romannumeral2}), as shown in Figs. \ref{fig:RamanSwitch}(a) and \ref{fig:RamanSwitch}(b). One can see that $P_1$ continues to increase with $I_0$, but the intracavity pump power, $P_0$, is clamped at ${\gamma_{1}}/2{g_{01}}$.  Thus, Mode \uppercase\expandafter{\romannumeral2} cannot derive sufficient gain solely from the pump field to start lasing. However, as shown in Fig. \ref{fig:RamanSwitch} (b), Mode \uppercase\expandafter{\romannumeral1} can also provide a gain mechanism for Mode \uppercase\expandafter{\romannumeral2} proportional to $P_1$, the second mode can be excited and the system undergoes a transcritical bifurcation when \begin{eqnarray}
I_{0,1}=\frac{\gamma_{1}}{8\gamma_{ex,0}g_{01}}[\gamma_{0}+\frac{1}{g_{12}}\frac{\omega_0}{\omega_1}(g_{01}\gamma_2-g_{02}\gamma_1)]^2,
\end{eqnarray} which can be obtained by taking $P_2=0$ and $-\gamma_2/2+g_{02}P_0+g_{12}P_1=0$ into Eq.(2)-(4).

As a consequence of the appearance of Mode \uppercase\expandafter{\romannumeral2} (regime \uppercase\expandafter{\romannumeral3}), Mode \uppercase\expandafter{\romannumeral1} is gradually suppressed, since the former opens a loss path for the latter through SRS \cite{boyd2003nonlinear}, with the loss being proportional to $P_2$. When two Raman modes lase simultaneously, $P_1$ and $P_2$ are determined from a simple linear relationship
\begin{equation}
g_{01}P_1+g_{02}\frac{\omega_1}{\omega_2}P_2=\frac{(g_{01}\gamma_{2}-g_{02}\gamma_{1})}{2g_{12}}.
\label{eq:E1E2condition}
\end{equation}
Clearly, both modes are still influenced by $I_0$. In the two-mode lasing regime, the intracavity pump power, $P_0$, increases with  $I_0$, while $P_1$  reduces with $I_0$ until 
\begin{eqnarray}
I_{0,2}=\frac{\gamma_{2}}{8\gamma_{ex,0}g_{02}}[\gamma_0+\frac{1}{g_{12}}\frac{\omega_0}{\omega_1}(g_{01}\gamma_2-g_{02}\gamma_1)]^2,
\end{eqnarray}
where  Mode \uppercase\expandafter{\romannumeral1} is completely switched off, see Figures \ref{fig:RamanSwitch}(a) and (b). Even with further increasing of the pump, Mode \uppercase\expandafter{\romannumeral1} cannot be turned on again and only Mode \uppercase\expandafter{\romannumeral2} remains on (regime IV). This counter-intuitive phenomenon can be explained by considering the following two aspects: Firstly,
the existence of Mode \uppercase\expandafter{\romannumeral2} reduces the Q-factor of Mode \uppercase\expandafter{\romannumeral1}, and the threshold of Mode \uppercase\expandafter{\romannumeral1} increases with $P_2$; Secondly, the gain provided to Mode \uppercase\expandafter{\romannumeral1} is fixed because Mode \uppercase\expandafter{\romannumeral2} causes the intracavity  pump power to be clamped. The latter 
has no correspondence with  conventional cascaded single-mode Raman lasers. Therefore, the mode switching induced by CLS cannot be readily illustrated as a unidirectional energy transfer between two lasing modes; in fact, it implies that the weak mode interaction could modulate the lasing dynamics of multimode lasers dramatically.

It is worth mentioning that there is no hysteresis phenomenon when the pump power, $I_0$,  is ramped down;  hence, it is possible to control the lasing modes simply by changing $I_0$.  It is convenient to control the intracavity power, $P_0$, by changing the relative detuning of the pump and cavity modes, therefore we perform the numerical calculation for the scanning pump case, as shown in Fig. \ref{fig:RamanSwitch}(c). It turns out that the four regimes can be achieved by controlling the detuning. As the detuning decreases, Mode \uppercase\expandafter{\romannumeral1} is excited first. Further detuning simultaneously turns on Mode \uppercase\expandafter{\romannumeral2} and suppresses Mode \uppercase\expandafter{\romannumeral1}. When the system is operated close to resonance, Mode \uppercase\expandafter{\romannumeral1} is annihilated completely and Mode \uppercase\expandafter{\romannumeral2} keeps growing until maximum coupling is reached.

\begin{figure}[h]
\centering
\includegraphics[width=\linewidth]{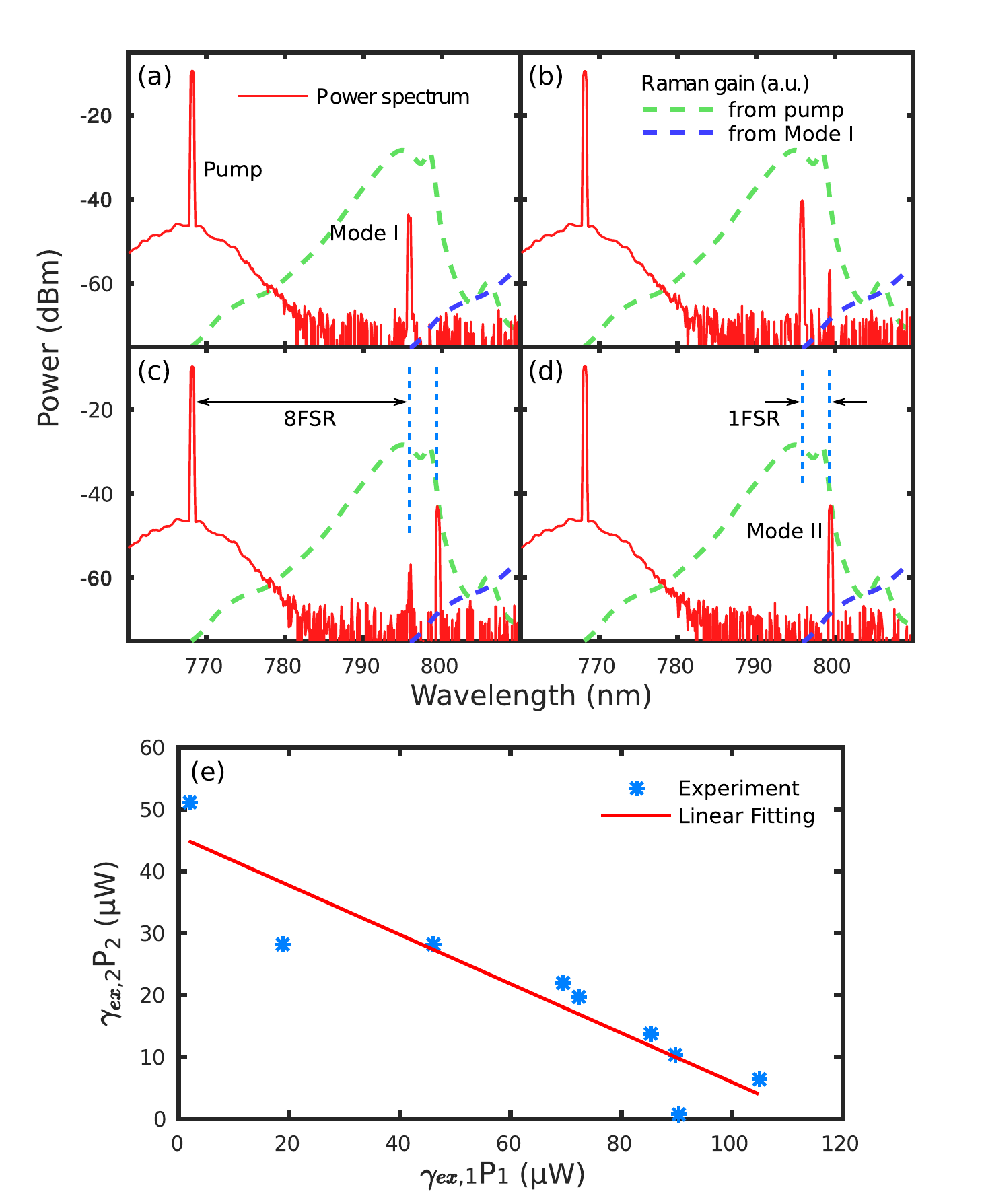}
\caption{Switching process of the Raman lines is observed by scanning the pump laser in the blue-detuned region of a cavity mode at 768.1 nm. From (a) to (d), the detuning of the pump is approximately set at 20 MHz, 18 MHz, 15 MHz and 10 MHz, respectively. (a) The first Raman laser at 795.9 nm appears (Mode \uppercase\expandafter{\romannumeral1}). (b) The second Raman line (Mode \uppercase\expandafter{\romannumeral2}) at 799.4 nm  appears. (c) Mode \uppercase\expandafter{\romannumeral1} is suppressed while Mode \uppercase\expandafter{\romannumeral2} becomes stronger. (d) Mode \uppercase\expandafter{\romannumeral1} is annihilated and only Mode \uppercase\expandafter{\romannumeral2} remains.  (e) The output powers of Mode \uppercase\expandafter{\romannumeral1} and Mode \uppercase\expandafter{\romannumeral2} are  negatively and linearly correlated with each other.}
\label{fig:ExperimentalPlot}
\end{figure}

To experimentally confirm the mode switching, we performed measurements using a silica microsphere with a diameter of around 41.5 $\mu$m, fabricated from a standard optical fiber reflowed by a $\rm CO_2$ laser \cite{murphy2017all}. A tapered optical fiber was used to couple the pump laser into the cavity through evanescent coupling, see Fig. \ref{fig:RamanDiag}(b). The output laser was divided into two paths, one for observing the transmission spectrum of the WGM resonator and the other for observing the Raman lasing spectrum. The transmission spectrum was recorded using a photodiode connected to a digital oscilloscope, and was used for locking the pump to the WGM. The Raman lasing spectrum was measured by an optical spectrum analyzer (OSA). To avoid  parametric oscillation and the coherent anti-Stokes Raman scattering, the pump wavelength was set to 768.1 nm - this corresponds to the normal dispersion regime for silica \cite{farnesi2014stimulated,agha2007four}. The input laser power was approximately  500 $\mu$W, and the laser frequency was finely tuned and thermally locked  to a cavity mode \cite{Carmon2004} with a Q factor of 9.7$\times 10^7$. During the thermal locking process, the maximum frequency change of the pump was less than 1 GHz, which of course would shift the Raman gain profile, however, this shift is negligible compared to the 1.65 THz frequency interval between the two lasing modes, as shown in Fig. \ref{fig:ExperimentalPlot}.  

The experimental results are presented in Fig. \ref{fig:ExperimentalPlot}.
As the laser frequency approaches resonance,  the first Raman lasing mode (Mode \uppercase\expandafter{\romannumeral1})  at 795.9 nm is excited, with its power increasing as the detuning decreases until the second Raman lasing mode (Mode \uppercase\expandafter{\romannumeral2}) at 799.4 nm appears. With a further reduction of the detuning, the power switching between the two Raman modes is observed, and, eventually, only Mode \uppercase\expandafter{\romannumeral2} remains on, as evidenced in Fig. \ref{fig:ExperimentalPlot}(d). The output powers of both Mode \uppercase\expandafter{\romannumeral1} and Mode \uppercase\expandafter{\romannumeral2} during the switching process were measured and are plotted in Fig. \ref{fig:ExperimentalPlot}(e). 

The powers of these two lasing modes are negatively and linearly correlated with each other during the switching process, in qualitative agreement with the theoretical model, see in Eq. \ref{eq:E1E2condition}. Note that the frequency spacings between the pump and two lasing modes are exactly integer numbers of the free spectral range (FSR), i.e., they belong to the same mode family and, therefore, have near unity overlap. Otherwise, the overlap coefficients would need to be introduced into Eq. (1). The mode overlap is particularly important for a standing wave resonator in which the Raman gain saturation could intrinsically lead to stable, single-mode lasing at a high power \cite{lux2016intrinsically}.  This is because the CLS may be suppressed due to weak mode overlap . In this scenario, the mode overlap might be two small to make CLS happen.

In these experiments, we selected resonators with a low number of high Q modes to avoid an overly complicated WGM spectrum and to reduce the chance of cascaded lasing beyond the first order. When we used an even higher pump power, or a smaller detuning, many high order cascaded lasing modes (up to at least 11 modes) were often observed. 
In some samples, 
we also observed the conventional cascaded lasing modes in single-mode fashion before the appearance of the second Raman lasing modes in the first order. This can be attributed to the differences in the Q factor and the FSR for different resonators.  There is an abundance of switching behaviors due to CLS when multiple lasing modes are involved; this is especially true when the phase-matching condition is satisfied and  other nonlinear optical phenomena occur simultaneously. Detailed analysis of this sophisticated process is beyond the scope of this work. 

In summary, although a lot of interest has been garnered, and many applications have been demonstrated, few studies have focused on understanding the details of cascaded light scattering in Raman lasers. In this work, we show that, aside from the application for extending the frequency range, naturally occurring CLS has a significant impact on the generation of Raman lasing  and cannot be ignored when multiple modes are involved. Subsequently, it may have an impact on the realization of Kerr-frequency combs \cite{Suzuki2018,liang2010passively,okawachi2017competition,kato2017transverse,cherenkov2017raman}, phase-locked Raman lasers \cite{liang2010passively,lin2016phase}, and soliton generation \cite{herr2014temporal,karpov2016raman,yang2017stokes}. Besides Raman lasers, this SRS interaction also exists in other lasers, such as rare earth doped microlasers, therefore, the mode switching induced by SRS  could provide a strategy to achieve a wavelength-switchable laser via CLS in a variety of laser systems.  Exploring this dynamically related phenomenon  \cite{zhukovsky2007switchable,lei2017pump} may find direct applications in all-optical, flip-flop memories \cite{liu2010ultra} and switchable light sources. 

 S. Kasumie and F. Lei contributed this work equally. This work was funded by the
Okinawa Institute of Science and Technology
Graduate University (OIST).


\end{document}